# Social Media for Emergency Rescue: An Analysis of Rescue Requests on Twitter during Hurricane Harvey

Lei Zou[1*], Danqing Liao[2], Nina S.N. Lam[3], Michelle Meyer[2], Nasir Gharaibeh[4], Heng Cai[1], Bing Zhou[1], Dongying Li[2]

[1]Department of Geography, Texas A&M University
[2]Department of Landscape Architecture and Urban Planning, Texas A&M University
[3]Department of Environmental Sciences, Louisiana State University
[4]Department of Civil Engineering, Texas A&M University

[*]Corresponding author: Lei Zou, lzou@tamu.edu

**Abstract:**

Social media plays increasingly significant roles in disaster response, but effectively leveraging social media for rescue is challenging. This study analyzed rescue requests on Twitter during the 2017 Hurricane Harvey, in which many residents resorted to social media to call for help. The objectives include: (1) understand the characteristics of rescue-request messages; (2) reveal the spatial-temporal patterns of rescue requests; (3) determine the social-geographical conditions of communities needing rescue; and (4) identify the challenges of using social media for rescue and propose improvement strategies. About half of rescue requests either did not provide sufficient information or neglected to include rescue-related hashtags or accounts. Of the 824 geocoded unique rescue requests, 41% were from FEMA-defined minimal flood risk zones. Communities sending more rescue requests on Twitter were environmentally and socioeconomically more vulnerable. Finally, we derived a framework summarizing the steps and strategies needed to improve social media use for rescue operations.

## 1 Introduction

Social media such as Twitter and Facebook create new channels to observe and manage social risk perceptions, communications, and behaviors during disasters (Houston et al. 2015; Shan et al. 2019). In hazard events, social media users can share their concerns, needs, opinions, and observations, and receive disaster-related information posted by official agencies anytime at any place (Wang and Ye 2018). At the same time, management organizations can use social media to interact with the public directly to obtain near-real-time human-centric information which are difficult to derive from traditional databases, such as people's reactions and disasters' societal impacts (Wang and Ye 2019). Therefore, many researchers have explored social media uses to improve disaster management, including early warning (Wu and Cui 2018), emergency rescue (Wang Z et al. 2020), damage assessment (Kryvasheyeu et al. 2016; Zou et al. 2018), recovery monitoring (Jamali et al. 2019), and resilience estimation (Dufty 2012; Wang et al. 2021).

During 2017 Hurricane Harvey, many coastal Texas residents resorted to social media to ask for help, marking Harvey as one of the first events in which social media played significant roles in fast-response and rescue missions. After Harvey made the first landfall in the United States on August 25, 2017, heavy rain barreled down on Houston throughout the weekend, causing devastating floods to local communities. Of the 10,000 Houston residents who failed to evacuate before the flood but needed rescue, only 3,000 were rescued by the local police or fire departments (Gallagher 2017). When the 911 system was overloaded and could not be connected, many victims turned to social media for help. Users posted rescue requests along with their addresses on social media in hopes of getting help from first responders or volunteers.

However, effectively employing social media in emergency rescue operations remains challenging due to the miscommunication between victims and disaster responders on social media (Mihunov et al. 2020). Many research questions need to be answered to enhance the efficiency and reliability of social media use for rescue in future events. For example, when, where, how, and by whom was social media being used for requesting rescue during disasters? What were the obstacles to using social media for emergency rescue? How to improve social media use for life-saving operations in future emergencies?

This study analyzed the rescue-request messages on Twitter during Hurricane Harvey from its first landfall in Texas (August 25, 2017) to the date it weakened to a tropical storm (August 31, 2017) to address the above questions. The objectives are four-fold: (1) to understand the characteristics of rescue-request messages on Twitter; (2) to reveal the spatial-temporal patterns of rescue requests during Harvey; (3) to determine the underlying geographical and social conditions of communities needing rescue; and (4) to identify the challenges of using social media for online rescue and propose improvement suggestions. First, the rescue-request tweets during Hurricane Harvey were collected using multi-criteria filters and manual labeling. We analyzed the information completeness and frequently used keywords, accounts, and hashtags of the rescue-request messages. Second, a geoparsing framework was developed to recognize and geocode locational information in Twitter messages for identifying communities that failed to evacuate and needed additional emergency resources. Third, we analyzed the geographical and socioeconomic conditions of the rescue-request communities through correlation analysis. Finally, we summarized the difficulties of using social media for rescue into a framework and proposed improvement strategies. Results from this study will shed light on disaster mitigation, preparedness, and rescue operations in coastal Texas and other hurricane-prone areas.

## 2 Background

### *2.1 Social Media Uses for Disaster Management*

Social media uses have penetrated every sector of human society, including disaster management. To efficiently explore the use of social media data and platforms during disasters, the first and foremost step is to develop algorithms extracting disaster-related information from the massive and noisy social media data. Verma et al. (2011) combined manual-annotation with automatically derived linguistic features for detecting situational awareness from Twitter during emergencies. Their method was able to correctly categorize 80% of Twitter messages as disaster-related or unrelated. A keyword-based text filtering algorithm was developed to detect and visualize disaster-related Twitter use frequencies for multiple types of natural hazards (Maldonado et al. 2016). Their algorithm accurately classified 93% of Twitter data as relevant to volcanos, earthquakes, weather events, fires, or others. Huang et al. (2018) proposed a visual-textual fused approach to tag flood-related Twitter messages automatically. Their results demonstrated that considering both text contents and images could significantly increase the precision of extracting flood messages compared with keyword-based approaches.

Meanwhile, many scholars have examined the diverse applications of social media in different phases of the disaster management cycle - preparedness, response, recovery, and mitigation. Sutton et al. (2014 & 2015) concentrated on reinforcing the content of emergency updates published by official organizations to alert more residents and motivate them to take action in the preparedness phase. Houston et al. (2015) developed a functional framework to facilitate the creation of disaster tools and formulate disaster management implementation processes for social media use in disaster responses. They suggested that social media can help governments and communities prepare and receive disaster warning information, signal and detect disasters prior to an event, and reconnect community members post a disaster. Li et al. (2018) proposed a novel algorithm using Twitter data to map flooded areas rapidly. Their model could visualize the flood extending in near real-time by fusing Twitter data with field gauge and elevation data. Another recent study developed a machine learning-based algorithm to identify people who experienced Hurricane Sandy disaster and assess their concerns by analyzing their Twitter data in the post-disaster recovery phase of Hurricane Sandy (Jamali et al. 2019). Their results reveal that information derived from mining social media could improve the understanding of priorities of people impacted by natural disasters, which is vital for effective recovery policymaking. Several studies have examined the value of social media data in post-disaster damage estimation (Guan and Chen 2014; Kryvasheyeu et al. 2016; Zou et al. 2018). Using Hurricane Sandy as an example, these studies found significant correlations between disaster-related Twitter activities and economic losses at multiple spatial and temporal scales.

The relationships between social media activities and disaster vulnerability and resilience were tested in previous investigations. A study examined the correlation between Twitter responses to Hurricane Sandy and community vulnerability in the northeastern U.S. (Wang et al. 2019). Their results show that physically vulnerable communities had more intense social responses, while socially vulnerable communities were digitally left behind with less disaster-related Twitter activities. Wang and others (2021) elaborated on social media use in understanding social and geographical disparities of disaster resilience. Their results indicate that county-level communities with higher disaster-related Twitter use during 2012 Hurricane Isaac were generally communities with better social-environmental conditions, implying the digital divide on social media use may exacerbate community disaster resilience inequalities.

*2.2 Hurricane Harvey: The U.S.'s First Social Media Storm*

Hurricane Harvey was formed on August 17, 2017 and made the first landfall in the United States at San Jose Island, Texas, on August 25, as a category-4 hurricane (Figure 1). Then Harvey moved inland and stalled at southwestern Houston before returning to the Gulf of Mexico on August 28. Two days later, Harvey made its final landfall in Cameron, Louisiana, on August 30, and quickly weakened and dissipated as it drifted inland. Harvey caused $125 billion damages, tied with 2005 Hurricane Katrina as the costliest tropical cyclones on record. Additionally, 68 direct and 35 indirect fatalities, more than 30,000 displacements, and more than 17,000 rescues were caused by Hurricane Harvey (Blake and Zelinsky 2018).

In Houston and its surrounding region, Harvey produced unprecedented precipitation, leading to historic-level flooding. As a result, many people were caught off-guard and needed emergency assistance to evacuate (Blake and Zelinsky 2018). The Houston 911 system, which usually handles about 8,000 daily calls, was overloaded by receiving more than 56,000 calls within 15 hours between August 26 to 27, 2017 (Gomez 2017). When Houston residents were unable to get through to the 911 system, they turned to social media to ask for help. Hashtags like *#sosHarvey* were used to flag citizen victims on social media, including Twitter, Instagram, and Facebook. Meanwhile, accounts like *@HarveyRescue* collected addresses of people who needed assistance and shared their information publicly. As one of the first events in the U.S. that social media played significant roles in disaster rescue, Harvey was referred as "the U.S.'s first social media storm" (Rhodan 2017).

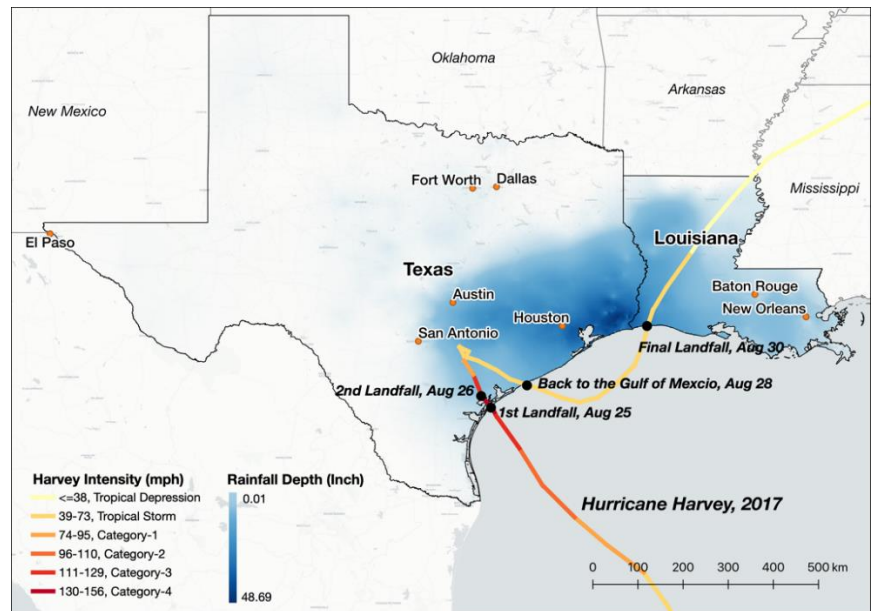

**Figure 1**. The track and precipitation of 2017 Hurricane Harvey

Several scholars have explored the emerging use of social media in Hurricane Harvey. Zou and others (2019) developed a Twitter data mining framework to calculate the public awareness and sentiment of residents during Hurricane Harvey. Their results indicate that many Houston residents were underprepared for the flooding. Disaster-related Twitter activities increased significantly two days after Harvey's first landfall when Houston residents were flooded and needed rescue. Yang and others (2017) built a text classifier to detect victims calling for help and volunteers offering rescue services during Harvey using the support vector machine (SVM) algorithm. A survey of 195 Twitter users who asked for rescue on Twitter during Harvey shows that 91% of users found Twitter very or extremely useful for flood rescue (Mihunov et al. 2020). Wang, Hu, and Jesoph (2020) developed a neuro-net toponym recognition (NeuroTPR) model for extracting locations from social media messages and tested the model to derive addresses from rescue requests on Twitter during Hurricane Harvey. Their results show that the NeroTPR model outperformed traditional name-entity-recognition (NER) tools in recognizing many fine-grained toponyms in rescue messages. These pioneered investigations have offered valuable information on social media use for disaster management and emergency responses during Hurricane Harvey.

The preceding examples also identify the challenges in using social media for disaster rescue (Mihunov et al. 2020). First, there was no official methods for emergency rescue requesting. Consequently, people composed their messages differently to ask for help, making it challenging to search for and locate disaster victims. Second, volunteers, who emerged and organized quickly during the storm, manually searched, read, and processed the large amount of social media data to find rescue-related messages, which required intensive human resources and time. Rapid and automated social media data mining and visualization tools are needed. Third, there was a miscommunication on social media between users and responders or volunteers. People who asked for help on social media did not know if assistance would arrive or not. More research on how, when, and where people used social media for relief in the past events is essential to identify the existing challenges, detect vulnerable communities, and inform improvements of using social media for rapid responses under emergencies.

## 3 Methodology

### 3.1 Twitter Data Collection and Cleaning

We chose Twitter as the social media data source for this study because Twitter is demonstrated as one of the primary tools to send and receive online rescue requests during Harvey (Mihunov et al. 2020). Twitter is a platform where users can post, share, or repost messages of 140 characters, refer to as tweets. Twitter has extended the character limit to 280 since Nov 2017, which was after Hurricane Harvey. We purchased the Twitter data from GNIP, a social media data aggregation company that provides full Twitter data since 2006 and was acquired by Twitter in April 2014. Harvey-related Twitter data were collected during August 17 - September 7, 2017, from the day when Harvey was named to two weeks after Harvey dissipated. We used a list of case-insensitive keywords about the disaster and related to known rescue agencies and volunteer groups to identify an initial collection of Harvey-related tweets: [*harvey, hurricane, storm, flood, houston, txtf* (Texas Task Force)*, coast guard, uscg* (U.S. Coast Guard)*, houstonpolice, cajun navy, fema* (Federal Emergency Management Agency)*, rescue*]. Every tweet containing at least one of the keywords was retrieved, resulting in a total of 47 million tweets in the initial collection. The obtained Twitter data were encoded in JavaScript Object Notation (JSON) format with 15 fields, including tweet id, user's profile, geo-tag, tweet content, timestamp, etc. This research used two attributes: text content ('body') and time when the tweet was created ('created_at').

Figure 2 illustrates the workflow of Twitter data collection, processing, and analysis in this study. The first step is selecting potential rescue-request tweets. Original English tweets containing five-digit numbers beginning with '77' in the initial GNIP Twitter database were chosen as potential rescue-request tweets for three reasons. First, we selected original tweets and removed retweets (reposted tweets) to avoid extracting duplicate rescue requests. Second, we assume that users were inclined to provide full addresses with zip codes in tweet contents when requesting help online so that they were more likely to be located and rescued. Third, this study focused on Houston and Beaumont-Port Arthur Metropolitan Statistical Areas (MSA), in which the zip codes are five-digit numbers starting with '77'. Although this method might overlook rescue-request tweets in non-English languages, outside of the study area, or with no zip code, the dataset generated from this approach could be used as a sample to understand online rescue-request behaviors and their spatial-temporal patterns. Furthermore, we could use this dataset to train machine learning algorithms to recognize rescue requests from the full database.

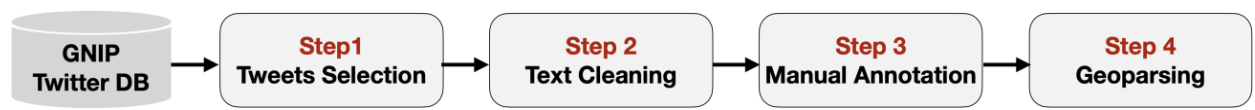

**Figure 2**. The workflow of Twitter data processing

The second step is text cleaning. We kept the '@' and '#' symbols and removed unrelated elements in each tweet, e.g., emoji, images, links, videos, and other special characters. In Twitter, the at-sign ('@') is used to mention or reply to other users, while the hashtag-sign ('#') can be used to create or search for tweets containing topics of interest. During Hurricane Harvey, users asking for rescue assistance created multiple hashtags (e.g., *#sosharvey*, *#harveyrescue*) to make their tweets searchable by disaster responders or volunteers. They also mentioned different accounts like the Houston Police Department (@houstonpolice) or Federal Emergency Management Agency (@fema) in tweets. We kept the two symbols to investigate the most frequently used hashtags and the most mentioned accounts in the subsequent analysis.

## 3.2 Manual Annotation

To better understand the information conveyed by each tweet, we applied manual annotation in step 3 to label Twitter data based on four questions (Figure 2): Is the tweet asking for help? Does the tweet provide a full address? Does the tweet mention demographic or health-related information of people who need assistance, e.g., gender, age, and physical conditions? Does the tweet explain the reason for using Twitter for rescue? For each question, we labeled the tweet as one if the answer is yes and zero for no. The first two questions are used to select rescue-request tweets with specific geographic information so that we could investigate how and where people sought help on social media. Answers to the second and third questions indicate whether Twitter users provided sufficient information for disaster responders to locate them and provide needed assistance. The last question collects the information on users' obstacles in requesting rescue using conventional approaches, which is valuable for pinpointing limitations in the current emergency responding systems and developing improvement suggestions.

Table 1 lists five labeled tweets to exemplify the manual annotation procedures and results. The street names and numbers in all tweets were replaced by '*street_name_x*' and '*999*', respectively, to preserve users' privacy. In the first tweet, the user was in need of rescue and provided a complete address, but did not mention the victim's information or give reasons for using Twitter, so its labels are [1, 1, 0, 0]. The second and third tweets have the victims' information. The former provides a full address while the latter contains an incomplete address, so their labels are [1, 1, 1, 0] and [1, 0, 1, 0]. In addition to providing an address and the victim's information, the fourth tweet explained why using Twitter for rescue, the 911 lines were busy, so it was labeled [1, 1, 1, 1]. Instead of requesting rescue, the fifth tweet shared the location of a shelter in Orange City in Texas, so the label is [0, 1, 0, 0].

**Table 1**. Examples of manually labelled tweets

| Tweet | Help | Full Address | Victim | Why |
|---|---|---|---|---|
| PLEASE PLEASE IN NEED OF RESCUE BEEN STRANDED FOR HOURS *999 street_name_1* APT HOU TX 77026 | 1 | 1 | 0 | 0 |
| @KHOU My friend her roommate and their dogs in attic house flooding *999 street_name_2* Drive Dickinson TX 77539 | 1 | 1 | 1 | 0 |
| #HOUSTON RESCUE Let someone know man named John is STILL on his roof water rising rapidly *999 street_name_3* Dr | 1 | 0 | 1 | 0 |
| Lines are busy Send help to *999 street_name_4* Houston Tx 77078 adults, kids and infant #rescue #houston | 1 | 1 | 1 | 1 |
| Orange City Shelter is at West Orange Elementary School at *999 street_name_5* Dr Orange TX 77630 | 0 | 1 | 0 | 0 |

### *3.3 Geoparsing*

The fourth step is geoparsing rescue-request tweets. Geoparsing is the process of recognizing and converting free-text descriptions of places into unambiguous geographic identifiers, namely coordinates. Geoparsing includes two parts: toponym recognition and toponym resolution. Toponym recognition extracts the complete address information from the textual contents, and toponym resolution converts the text addresses to geographical coordinates. During the manual annotation of the tweets in step 3, we noticed that Twitter users described addresses irregularly in informal sentence structures with name abbreviations, and some descriptions were misspelled. Therefore, the performance of the traditional Name-Entity-Recognition (NER) approach in recognizing locations from tweets is limited (Wang J et al. 2020). A better geoparsing tool specifically designed for extracting locational information from rescue-request tweets is needed.

In the first part of geoparsing, we designed a rule-based method to recognize toponyms from the tweet contents. The formal description of an address in the United States is [*Street Number, Street Name, Apartment Number (optional), City, State, Zip Code*]. Since all rescue-request tweets in this study contain zip codes, the proposed method extracts the full address in each tweet by locating the zip code as the ending point and searching for the starting point based on the following conditional criteria. If there is only one number before the zip code, then the full address is the text from the first number to the zip code. If there are at least two numbers before the zip code, we examined if there are address-related keywords (apartment, drive, road, avenue and street, and their abbreviations) between the nearest and second nearest numbers before the zip code. If yes, the text from the second nearest number to the zip code was extracted as the full address. Otherwise, the content between the zip code and its closest number was considered as the full address.

The second part of geoparsing, toponym resolution, could be accomplished through geocoding service providers, e.g., Google, ESRI, OpenStreetMap, Texas A&M Geocoder, Mapbox, etc. A previous study (Zou et al. 2018) suggest that Google geocoding service can tolerate misspelled addresses and provide accurate toponym resolution results for tasks with small amounts of data (no more than 40,000 free requests per month by each Google developer account). Hence, we chose the Google Geocoding Application Programming Interface (API) to geo-reference addresses derived from tweets into coordinates. Repeated coordinates in the geocoding results were removed to avoid double-counting rescue requests from the same addresses in the subsequent spatial-temporal analysis. We further compared the geocoded zip codes and tweet-derived zip codes to validate the geocoding results and removed addresses with unmatched zip code information or outside of Texas.

### 3.4 Geographical and Socioeconomic Data

In addition to the Twitter data, we collected two geographical and ten socioeconomic variables at the block group level (Table 2) to examine the conditions of communities sending rescue requests on Twitter. The mean elevation in each block group was derived from the Shuttle Radar Topography Mission (SRTM) dataset provided by the U.S. Geological Survey (USGS). The data were acquired in 2000 at the spatial resolution of 30m*30m. The rainfall depth from Harvey was obtained from Zou et al. (2019). They collected rainfall observations during Harvey from the National Hurricane Center and interpolated the observations into a 30m*30m raster using the Empirical Bayesian Kriging method. We calculated the averaged elevation and averaged interpolated rainfall depth of all grids for each block group.

**Table 2**. List of geographical and socioeconomic variables

| Category | Variable | Source |
| --- | --- | --- |
| Geographical | Mean elevation within the block group, 2000 | USGS |
| | Averaged interpolation of rainfall depth from Harvey, 2017 | NOAA |
| Socioeconomic | Median Household Income, 2017 | U.S. Census |
| | % Labor force employed, 2017 | U.S. Census |
| | % 25 years old with a bachelor's or a higher degree, 2017 | U.S. Census |
| | % Population 65 years and over, 2017 | U.S. Census |
| | % Households with telephone services available, 2017 | U.S. Census |
| | % Owner-occupied households, 2017 | U.S. Census |
| | % Households with no vehicle, 2017 | U.S. Census |
| | % White population, 2017 | U.S. Census |
| | % African American population, 2017 | U.S. Census |
| | % Hispanic population, 2017 | U.S. Census |

We collected ten socioeconomic variables from the U.S. Census' 2013-2017 American Community Survey (ACS) 5-year estimates to calculate community vulnerability. The selection of these variables was based on three reasons. First, a synthesis review of 174 articles published from 2005 to 2017 on disaster vulnerability and resilience assessment identifies six most frequently used socioeconomic aspects, including income, employment, education, age, community capacity, and housing capital (Cai et al. 2018). We included them for the analysis to enable future comparison. Second, we added the percentage of households with no vehicle available because having access to working vehicles during hurricanes is vital for evacuation, and this variable was found to be significant in indicating vulnerability and resilience to coastal hazards in previous studies (Cutter et al. 2008; Cai et al. 2016; Kirby et al. 2019; Van Zandt et al. 2012; Wang et al. 2021). Finally, three demographic variables, including the percentages of white, African American, and Hispanic population, were included because race and ethnicity affect social media use behaviors (Zhai et al. 2020). Race and ethnicity are also strong predictors of evacuation capacity and timing (Bolin and Kurtz 2018; Meyer et al. 2018), and therefore impact community disaster resilience (Cutter et al. 2014).

*3.5 Methods of Analysis*

First, we investigated the textual characteristics of rescue-request tweets to understand how people composed social media messages to ask for help during Hurricane Harvey. Basic descriptive statistics were generated to examine whether users provided complete information when asking for assistance on Twitter. We further examined the most frequently used keywords, hashtags, and accounts in rescue-request tweets.

Second, we analyzed the spatial-temporal patterns of rescue requests on Twitter during Harvey through mapping the locations of the rescue requests by block groups. We compared the rescue requests with the FEMA-defined flood risk areas to evaluate the reliability of FEMA's flood zone products in predicting rescue needs and guiding flood management.

Third, we identified the geographical and socioeconomic conditions of communities having households needing rescue and compared them with the conditions of all block groups in the Houston and Beaumont-Port Arthur MSAs. Pearson's correlation analysis was conducted between percentages of rescue-request households and the selected variables to further unravel the characteristics of communities needing additional rescue resources during emergencies.

Finally, we concluded with a few challenges facing the use of Twitter in emergency rescue based on our study findings and suggested an operational framework to improve its efficiency.

**4 Results**

*4.1 Textural Characteristics of Rescue-Request Tweets*

During August 25-31, 2017, 3,062 unique English tweets were extracted from the 47 million Harvey-related tweets as potential rescue-request tweets, and 1,865 (60.91%) were labeled as seeking rescue assistance (Table 3). Of the 1,865 rescue-request tweets, 1,667 (89.48%) contain full addresses, 1,230 (66.02%) describe victims' information, and only 1,138 (61.08%) provide both information (Table 2). Our database reveals that 10.52% of people did not include detailed locations when requesting rescue, posing additional challenges to pair nearby first responders with them. Victims' demographic or physical conditions are also vital for arranging rescue plans. Information such as the number of people who need help and whether special needs are required for people with disabilities or health conditions could help first responders allocate rescue teams and prepare required equipment or medical supplies. However, around one-third of users neglected to provide such information in their rescue-request messages.

**Table 3**. Manual annotation summary of rescue-request tweets

| Potentially Rescue-Related Tweets | Count | Proportion |
|---|---:|---:|
| Total | 3,062 | - |
| Request Rescue | 1,865 | 100% |
| Rescue Request with Full Address | 1,667 | 89.38% |
| Rescue Request with Victim Information | 1,230 | 65.95% |
| Rescue Request with Full Address and Victim Information | 1,138 | 61.02% |
| Explain Reasons for Using Twitter for Rescue | 21 | 1.13% |

An in-depth review of the 21 tweets that discussed why they used Twitter for rescue during Harvey indicates that the major reason (11 out of 21) was their inability to connect to 911. Other reasons included receiving no help after calling 911, no phones available to use, and helping others to ask for rescue. This result is consistent with an online survey of Twitter users who sent tweets for help during Harvey (Mihunov et al. 2020). As discussed earlier, the Houston police department received about seven times more calls than usual during Harvey. They were unable to answer and help all residents who needed evacuation. Consequently, residents used social media as an alternative option to ask for rescue when traditional approaches did not work. This result suggests that Twitter or other social media platforms, if better organized, could be used as an effective communication channel for rescue requests during emergencies.

To understand how users composed their rescue-request tweets, we summarized the frequencies of words, hashtags, and accounts used in the 1,863 tweets. Figure 3 shows the top 20 words, including four location-based, eight rescue-related, and three relevant to victims' information. The Houston police and the Cajun Navy, representing the governmental and voluntary first responders, were also frequently mentioned in the rescue-request tweets. The Cajun Navy are informal volunteer groups first formed in the aftermath of 2005 Hurricane Katrina and reactivated during the 2016 Louisiana floods, gaining national attention (Morris 2016). Since then, the different Cajun Navy groups have rescued thousands of citizens during numerous flooding events including Hurricane Harvey.

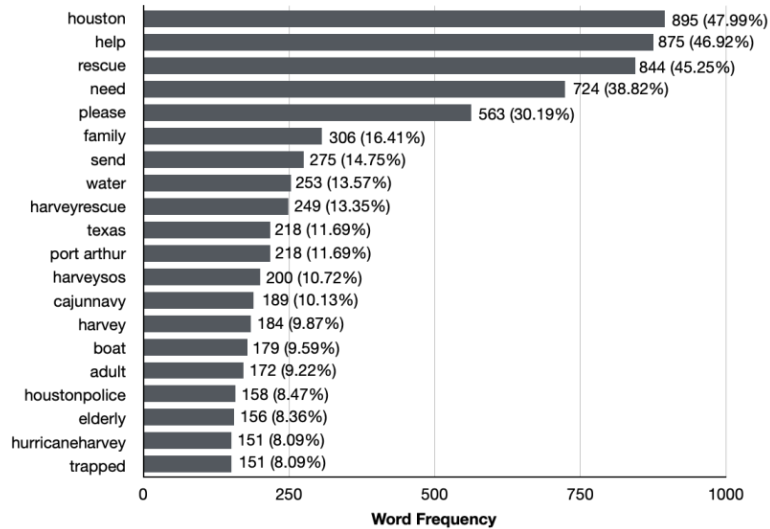

**Figure 3**. Top 20 keywords used in rescue-request tweets

Figure 4 summarizes the frequencies of hashtags and accounts in the rescue-request tweets. Hashtags are useful for emergency responding organizations and volunteers to search for rescue-request tweets. An examination of the number of attached hashtags in each tweet reveals that the majority (53.84%) of those rescue-request tweets do not contain any hashtag (Figure 4a). This result is consistent with an online survey of people who used Twitter for rescue during Harvey, in which 75% of the respondents did not use hashtags in their tweets asking for help (Mihunov et al. 2020). The ten most frequently used hashtags were *#harveysos*, *#cajunnavy*, *#harvey*, *#hurricaneharvey*, *#harveyrescue, #houston, #hosutonflood, #rescue, #help, and #houstonstrong*, which were attached in 10.31%, 9.02%, 8.05%, 7.51%, 7.46%, 5.85%, 4.94%, 4.35%, 2.31%, and 2.25% of tweets, respectively (Figure 4b). Among them, five are rescue-related (*#harveysos*, *#cajunnavy*, *#harveyrescue*, *#rescue*, *#help*), three are location-based *(#houston*, *#houstonflood*, *#houstonstrong*), and two are created by the event's name (*#harvey*, *#hurricaneharvey*). Hashtags based on the affected area or the event's name were not specific for rescue requests. Monitoring those hashtags could return tweets on any topics related to Hurricane Harvey or the city of Houston, most of which were unrelated to people needing help.

Mentioning the accounts of emergency responding organizations in tweets (e.g., police or fire departments) could help lead the messages directly to them and gain the attention of first responders. We examined the number of accounts mentioned in each rescue-request tweet and found that nearly half (45.84%) of them did not mention any account (Figure 4c). The most commonly cited accounts in the rescue-request tweets were the Houston Police Department (@houstonpolice), the ABC Television Station in Houston (@abc13houston), a voluntary group called to assist Harvey relief (@harveyrescue), CBS-affiliated television station in Houston (@khou), the U.S. Coast Guard (@uscg), a volunteer organization (@altnoods), the American Red Cross (@redcross), NBC-affiliated television station licensed to Houston (@kprc2), the Cable News Networks television channel (@cnn), and the U.S. National Guard (@usnationalguard). The top ten accounts mentioned in the rescue-request tweets included three governmental emergency responding agencies (@houstonpolice, @uscg, and @usnationalguard), four national and local broadcasting stations (@abc13houston, @khou, @kprc2, and @cnn), one non-profit charities

(@redcross), and two volunteer groups supporting rescue operations during Hurricane Harvey (@altnoods, @harveyrescue).

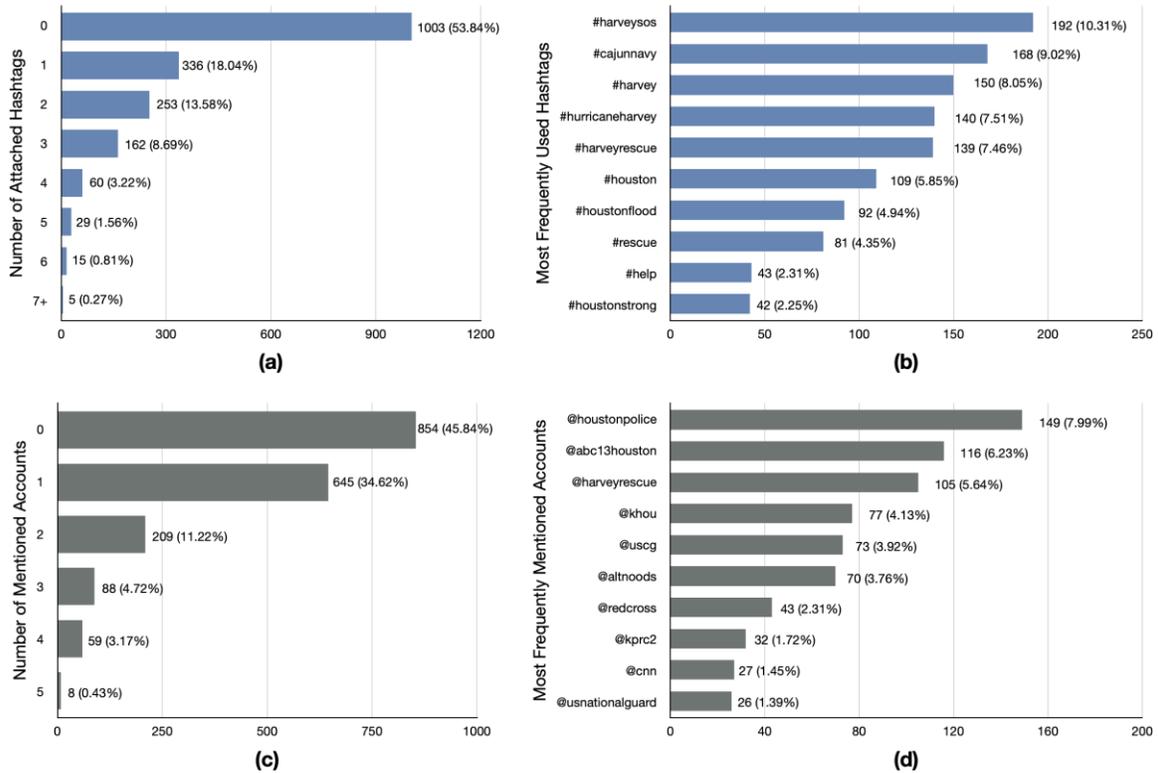

**Figure 4**. Frequency analysis of hashtags and accounts in the rescue-request tweets: (a) top 10 most frequently used hashtags; (b) frequencies of numbers of attached hashtags; (c) top 10 most frequently used accounts; (d) frequencies of numbers of mentioned accounts.

## *4.2 Spatial-Temporal Patterns of Rescue Requests*

A total of 1,031 (61.85%) unique addresses were extracted from the 1,667 rescue-request tweets with full addresses, and 1,028 (61.67%) were successfully geocoded through the designed geoparsing method. We further removed duplicated and zip-code mismatched records, resulting in 824 pairs of coordinates for the spatial-temporal analysis (Table 4).

Table 4. Geoparsing results from rescue-request tweets

| **Rescue-Request Tweets** | **Count** | **Proportion** |
| --- | --- | --- |
| All rescue-request tweets with full addresses | 1,667 | 100% |
| Step 1: Duplicate textual addresses removed | 1,031 | 61.85% |
| Step 2: Successfully geocoded | 1,028 | 61.67% |
| Step 3: Duplicate coordinates removed | 832 | 49.91% |
| Step 4: Addresses with mismatching zip code removed | 824 | 49.43% |

Figure 5 and Table 5 show the spatial-temporal patterns of rescue requests during Harvey in the Houston and the Beaumont-Port Arthur MSAs. The majority of the 824 unique rescue requests were sent on August 27th (212, 25.73%) and 28th (378, 45.87%) from the Houston MSA, while most of the rescue requests in the Beaumont-Port Arthur MSA were sent on August 30th. The rescue request locations are distributed unevenly in the two MSAs with four hotspots based on the point density. Hotspot A is the residential area located in east Houston between Interstate Highway 610 and Texas State Highway 8. Hotspot B is the Bellaire community near the southwest of Interstate Highway 610. Hotspot C is the Clodine town located between Texas State Highways 6 and 99 in western Houston. The last one is in the city of Port Arthur near Texas State Highway 73 along the shoreline of Sabine Lake (D).

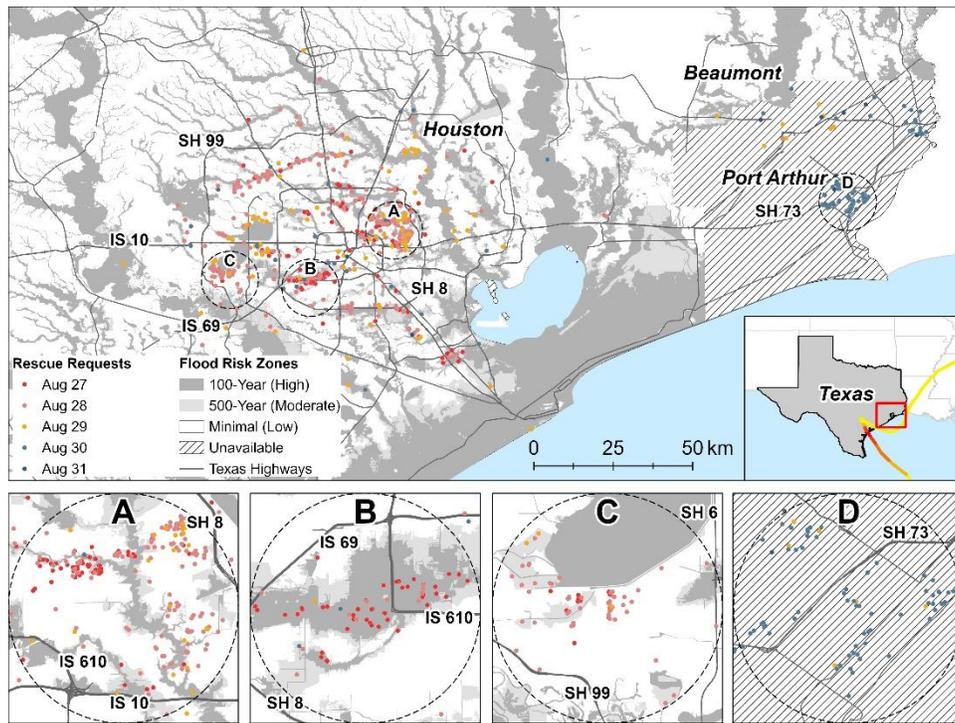

**Figure 5**. Locations of identified rescue requests from Twitter during 2017 Hurricane Harvey

Figure 5 and Table 4 also uncover the flood risks of locations sending rescue requests. Dark grey and light grey areas in Figure 5 are zones with a 1-percent and 0.2-percent annual chance of flood, referred to as the FEMA-defined 100-year (high risk) and 500-year (moderate risk) flood zones (FEMA 2020; Qiang et al. 2017). Among the 824 requests, 158 (19.18%) and 192 (23.30%) were in high and moderate flood risk areas, respectively, and 341 (41.38%) were sent from minimal flood risk zones. Specifically, although nearly all rescue requests in hotspot B were in high flood risk zones, most rescue requests in hotspots A and C were located in minimal flood risk areas. The available FEMA's flood hazard map does not provide data describing the flood risk levels in the Beaumont-Port Arthur MSA (shaded areas in Figure 5). The federal, state and local governments and residents have widely used FEMA-defined flood zones for disaster preparedness and response decision-making. During Hurricane Harvey – the costliest hurricane in U.S. history, however, over 40% of Houston residents needing life-saving assistance lived in areas outside of the 100/500-year flood zones, based on the Twitter-extracted information. This observation points out that people living in minimal flood risk areas also need to take actions in flood preparedness and response, e.g., early evacuation, to mitigate the flood threat to human safety, especially in catastrophes like Harvey.

**Table 5**. Temporal trends and proportions of rescue requests in different flood-risk-level zones

| Date | Flood Risk Levels | | | | Daily Total |
|---|---|---|---|---|---|
| | 100-Year (High) | 500-Year (Moderate) | Minimal (Low) | Unavailable | |
| 27-Aug | 85 | 42 | 85 | 0 | 212 (25.73%) |
| 28-Aug | 53 | 113 | 203 | 9 | 378 (45.87%) |
| 29-Aug | 16 | 30 | 41 | 20 | 107 (12.99%) |
| 30-Aug | 4 | 5 | 11 | 97 | 117 (14.19%) |
| 31-Aug | 0 | 2 | 1 | 7 | 10 (1.22%) |
| **Zonal Total** | 158 (19.18%) | 192 (23.30%) | 341 (41.38%) | 133 (16.14%) | 824 (100%) |

*4.3 Geographical and Socioeconomic Conditions*

We focused on the Houston and Beaumont-Port Arthur MSAs to determine the geographic and socioeconomic status of communities sending rescue requests on Twitter. We aggregated the rescue-request messages at the Census block group scale, the smallest geographical unit with a population of 600 to 3,000 people for which the Census Bureau publishes publicly available sample data.

The two MSAs consist of 3,333 block groups, and 374 (11.22%) of them contain rescue requests collected from Twitter during Harvey (Figure 6). The number of requests in these block groups ranged from 1 to 46, with an average value of 2.19. The percentages of households sending rescue requests on Twitter ranged from 0.01% to 4.19%, and the mean was 0.35%. The block group with the highest percentage is in Northeast Houston. It has 334 households, and 14 of them asked for help on Twitter during Harvey. We further classified those block groups with rescue requests based on z-score values of percentages of rescue-request households. The z-score measures how many standard deviations (SD) a value below or above the mean, which provides a more meaningful classification for categorizing variables (Cai et al. 2018). Two z-score values, -0.5 and 0.5, were chosen as the breakpoints, resulting in three categories, including low (0.01-0.11%), medium (0.11-0.58%), and high (0.58-4.19%). High-percentage block groups are mostly concentrated in the northeastern Houston MSA and southeastern Beaumont-Port Arthur MSA.

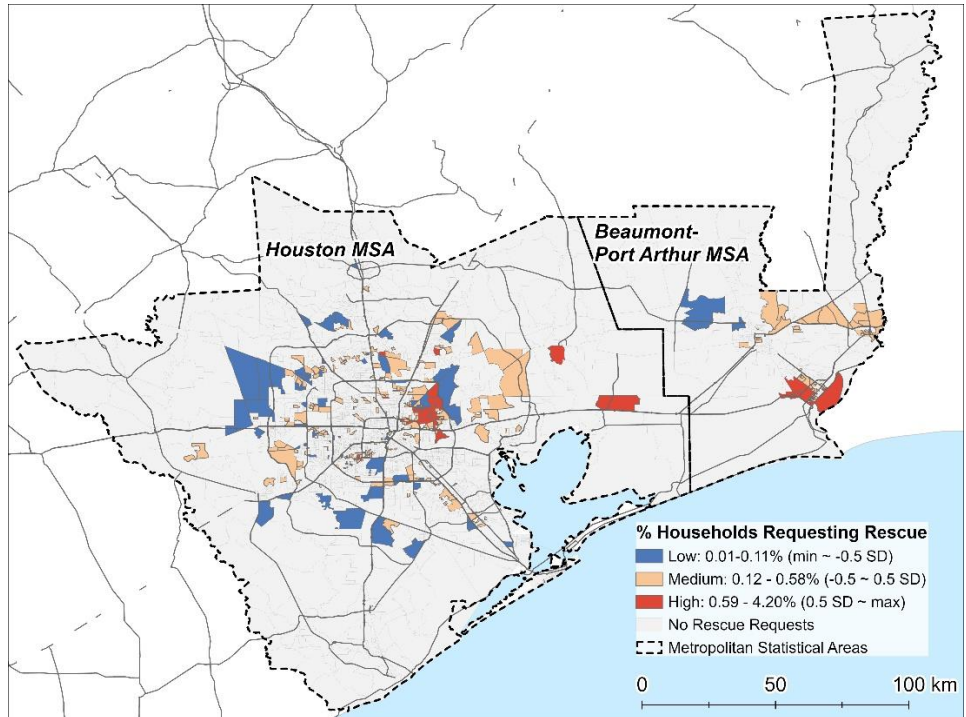

**Figure 6**. Block groups sending rescue requests on Twitter during Hurricane Harvey in the Houston and Beaumont-Port Arthur MSAs.

Figure 7 is the box plot of the 12 selected variables' value distributions in block groups of the three percentage categories and the whole study area, which visualizes the geographical and socioeconomic conditions of block groups with rescue-request tweets. In terms of the geographical factors, the elevations in over 75% of the high-percentage block groups are significantly below the median elevation at the block group level in the study area. On the other hand, most high-percentage and medium-percentage block groups suffered more rainfall from Harvey among all block groups in the two MSAs. The result demonstrated that neighborhoods that sent more rescue requests on Twitter, implying those who needed additional emergency responding resources, were communities that were more physically affected by Hurricane Harvey.

Regarding the socioeconomic status, communities with a higher percentage of households requesting rescue on Twitter were more socially vulnerable—relatively lower median household income, smaller percentage of 25-year-olds with a bachelor's or higher degree, lower employment rate, and more households without access to a vehicle—compared to all block groups in the two MSAs. There are no significant disparities among the distributions of the four block-group categories in terms of the percent of 65-year-old residents, percent of telephone services available, and percent of owner-occupied households. This could be interpreted from the following possible reasons. First, although social media were found more popular among the younger population in a few early social media analysis literature (Li et al. 2013; Kent et al. 2013), social media use increased among all age groups and became ubiquitous in 2017. Therefore, using social media was no longer a privilege for the younger population. Social media users of different ages might try all available methods to ask for help in emergencies (Mihunov et al. 2020). Second, although the percent of households with telephone services was a significant vulnerability and resilience indicator in some previous studies (Cutter et al. 2010; Cutter et al. 2014), this indicator did not show much variation in the study area. Over 92% of households in all block groups had available telephone services (i.e., cell phones, landlines, or other phone devices). Meanwhile, using social media for rescue requests does not require a phone but the Internet. Third, others outside the area might make requests on social media on behalf of residents or relatives living in affected areas. Lastly, homeownership rates are usually related to communities' general economic vitality and recovery (Cutter et al. 2014), but whether homeowners or renters need additional assistance during emergencies is unclear and needs more investigations. Overall, this analysis suggests that most rescue requests were sent from communities that were more socioeconomically vulnerable to coastal hazards.

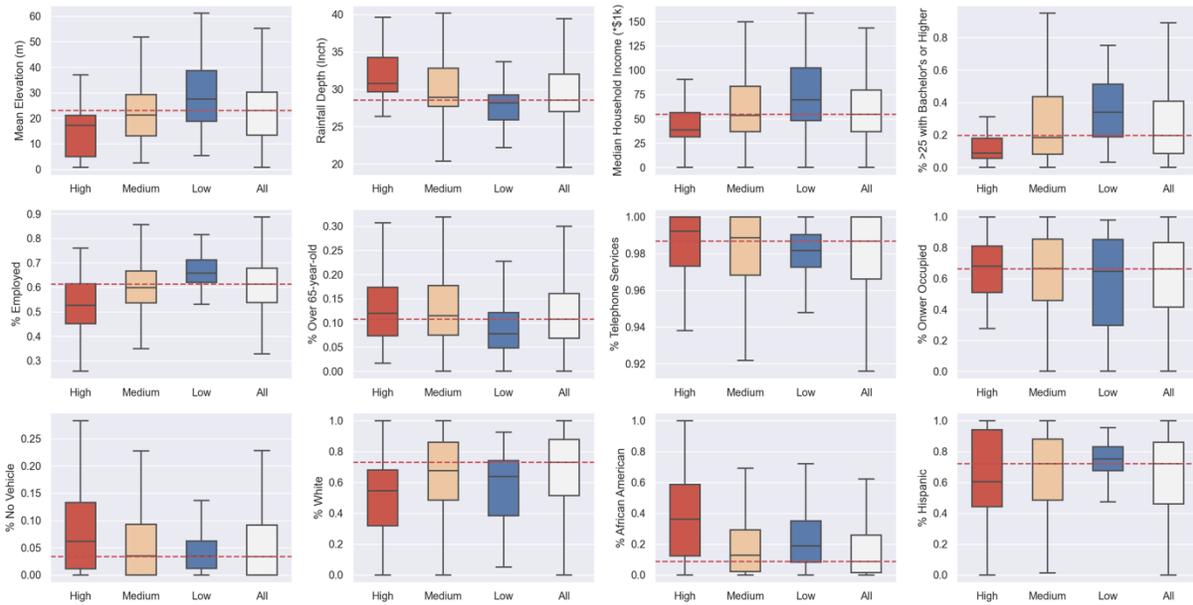

**Figure 7**. The distribution of geographical-social characteristics among block groups with different percentages of households sending rescue requests on Twitter during Harvey.

To further elucidate what communities having more people requesting rescue on social media during disasters, the correlations between the percentages of rescue-request households and the selected variables were examined. Block groups with very few requests on Twitter might not necessarily indicate that the whole community needed additional rescue resources. Therefore, we tested the correlations in three selections of block groups that had at least one, two, and three rescue requests on Twitter during Harvey, denoting as tests 1-3 in table 6. The same eight variables were found significant in all three tests. Factors positively correlated with the proportions of households requesting rescue on Twitter are rainfall depth, the percentage of 65-year-olds, and the percentage of African American population. Conversely, mean elevation, median household income, percentage of people over 25 with a bachelor's or higher degree, percentage of the employed population, and the percentage of the population are negatively correlated with the rescue-request household rates. The most significant factors in all three tests are the percent employed and the percent of African Americans. Both Figure 7 and Table 6 confirm that more households in environmentally and socially vulnerable communities sent rescue request on Twitter during Hurricane Harvey.

**Table 6.** Correlations between rescue-request household percentages and variables

| Variables | Percentage of Rescue-request Households | | |
|---|---|---|---|
| | **Test 1** (n = 374) # Requests ≥ 1 | **Test 2** (n = 123) # Requests ≥ 2 | **Test 3** (n = 64) # Requests ≥ 3 |
| Mean Elevation | -0.250** | -0.292** | -0.269* |
| Rainfall Depth | 0.219** | 0.272** | 0.273* |
| Median Household Income | -0.139** | -0.248** | -0.294* |
| % >25 with Bachelor's | -0.177** | -0.300** | -0.380** |
| % Employed | -0.317** | -0.391** | -0.471** |
| % Over 65-year-old | 0.164** | 0.188* | 0.279* |
| % Telephone Services | 0.066 | 0.094 | 0.068 |
| % Owner Occupied | 0.058 | 0.054 | 0.162 |
| % No Vehicle | 0.091 | 0.173 | 0.183 |
| % White | -0.184** | -0.286** | -0.351** |
| % African American | 0.279** | 0.381** | 0.446** |
| % Hispanic | -0.022 | -0.057 | -0.092 |

\*\* Correlation was significant at the 0.01 level (2-tailed). * Correlation was significant at the 0.05 level (2-tailed).

### 4.4 A Framework for Analyzing Emergency Rescue

Based on the above analysis, we identified three challenges of using social media for rescue during Hurricane Harvey and proposed a practical framework to address them (Figure 8).

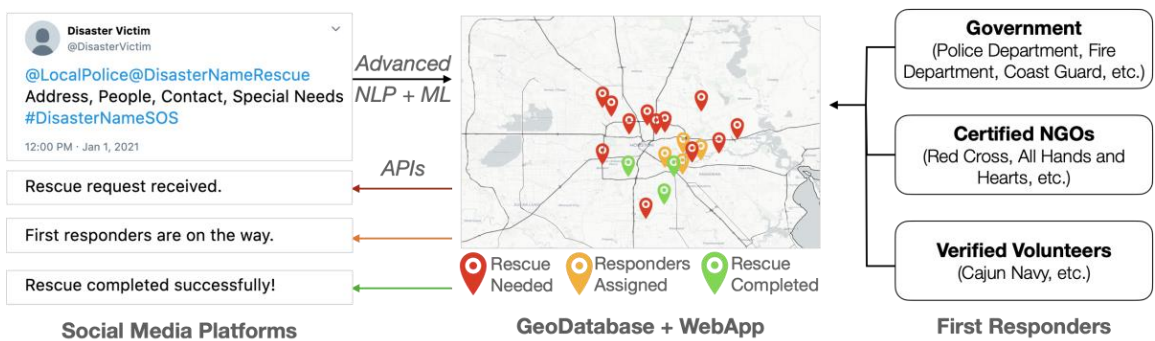

**Figure 8**. The proposed practical framework of using social media for emergency rescue.

First, people needing assistance did not compose the most effective rescue requests on social media platforms by including key information. The majority of the rescue-request tweets during Harvey did not include any hashtags. Although the rest contained hashtags, some popular hashtags are based on the general disaster names or affected locations. These hashtags may mean that time-sensitive rescue requests get lost in a deluge of all types of disaster information and commentary. Nearly 40% of the rescue-request tweets did not provide sufficient information either, e.g., addresses and special needs, which are vital for first responders to locate disaster victims and allocate rescue resources. Social media users in the hazard-prone area need to be informed in the disaster mitigation and preparedness phases about which hashtags and information should be included in the rescue-request messages. To address this challenge, we suggest emergency management agencies and nonprofit response organizations providing educational messaging pre-disaster on how to best compose social media messages requesting rescue and communicate their needs. Hashtags specifically for rescue requesting should be shared in the disaster mitigation and preparedness phases. Thus, using a hashtag collates requests for faster processing by rescuers. Some example hashtags are #DisasterNameRescue and #DisasterNameSOS.

Second, governmental agencies or volunteer organizations offered little information before Harvey on whether they would monitor social media activities and provide help. Nearly half of the rescue requests did not mention any organization's account, and the rest referred to numerous different organizations. As a result, the requests were placed into the void of social media timelines and may or may not be seen by rescuers. However, whether disaster victims should use social media to request rescue during emergencies is still controversial. Figure 8 shows the recommended help requesting methods posted by two of the top five most frequently mentioned accounts in rescue-request tweets. The non-governmental @HarveyRescue account encouraged people to ask for rescue on social media by including their complete addresses, the number of people needing rescue, phone number, special needs, and tag #HarveySOS and mention the HarveyRescue account in the tweet (Figure 9a). They also recommended users update rescue status as they were able. On the contrary, the U.S. Coast Guard's Twitter account suggested that users do not report their information on social media sites (Figure 9b), possibly due to the concerns of collecting unverified information and publicly sharing personal information. We suggest that governmental agencies, non-governmental organizations, and volunteers could open social media communication channels during large-scale emergencies and link the requests to first responders. Accounts of organizations providing emergency responses should be broadcasted on social media in the pre-disaster phase so that affected residents can tag them to ask for help when composing rescue requests.

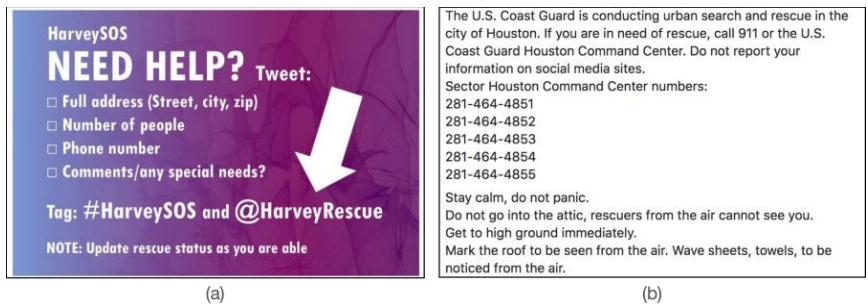

**Figure 9**. Examples of rescue-request methods shared on Twitter during Hurricane Harvey by (a) a non-governmental account @HarveyRescue, and (b) the U.S. Coast Guard.

Finally, harvesting rescue requests from big noisy social media data for life-saving operations are complicated in practice, partially due to the first two challenges. Since there is no standard approach for disaster victims to request rescue through social media platforms, there is no simple yet effective method for first responders to search those messages and extract vital information. Several key questions need to be addressed to leverage social media for emergency rescue effectively: how to develop tools to collect and process social media data automatically, recognize rescue-request messages, and extract needed information? Who should have access to social media-derived rescue requests? How to enable communication between disaster victims and first responders on social media? The recent advances in Natural Language Processing (NLP) and web-based geographic information systems bring opportunities to resolve these issues. First, models based on advanced NLP and Machine Learning (ML) algorithms can be developed to recognize social media messages asking for help and extract locational and other relevant information from those messages. Second, geodatabases and web applications can be used to store rescue requests and manage the life-saving operation. Authorized users, including governmental organizations (e.g., police and fire departments), certified non-governmental organizations (e.g., Red Cross), and verified volunteers, could log in to the web application and obtain victims' information to send help. Third, updates on rescue operations can be posted as replies to the original rescue-request tweets through Twitter APIs to avoid collecting duplicated rescue requests due to reposting the same requests on social networks.

## 5 Limitations and Future Research

This study has some limitations which could be addressed in future research. First, the dataset developed in this study does not contain all rescue-request tweets. Rescue requests missing zip codes, sending from outside of the study area, or posting by non-English Twitter users, were not considered in this investigation. This limitation could be improved by including translators in the manual analysis of tweets or applying auto-translation models or APIs to convert all rescue-related messages into English and leveraging advanced natural language processing and neural network algorithms to extract unstandardized addresses from those messages.

Second, the efficiency of rescue-request messages using different keywords and hashtags, contacting different accounts, and providing diverse information was not evaluated. Whether and how volunteers received various rescue-request tweets and how long it took them to find those tweets and send help are unknown. These questions could be answered by either tracking how different rescue requesting tweets were transmitted on social networks and received by first responders or conducting a questionnaire survey for people who sent rescue requests on Twitter to evaluate their messages' efficiency.

Third, not all vulnerable communities needing additional rescue assistance could be identified by mining Twitter activities. Social media use disparities exist across different social groups. The vulnerable population who requests help the most might not have access to social media platforms and asked for help for themselves online. Consequently, vulnerable communities with little social media use may not be identified from Twitter-derived rescue requests. This challenge could be resolved by searching communities having geographical and socioeconomic characteristics similar to or worse than those vulnerable communities reflected on Twitter.

Finally, rescue-related social media activities other than asking for help were not analyzed in this research. When manually labeling the potentially rescue-request tweets, we noticed that some non-rescue-request tweets are related to rescue activities. For instance, some tweets shared rescue-request methods, shelter locations, and the contact information of first responders. Such information is timely and critical under emergencies since it could indirectly help disaster victims seek assistance and evacuate. However, sharing those messages with the same hashtags or keywords used by rescue-request tweets, e.g., *#harveysos*, may pose additional difficulties for first responders to search rescue requests on social media. How such information spreads on social media and is received by people needing rescue and how to optimize the information sharing process necessitate further investigation.

## 6 Conclusion

This study examined the novel application of Twitter in emergency rescue during Hurricane Harvey in the Houston and Beaumont-Port Arthur MSAs. A total of 3,062 potential rescue-request tweets were extracted from the complete Twitter database for analysis. The study fulfilled the four proposed objectives. First, we summarized the characteristics of rescue-request messages on Twitter. A significant number of rescue-request tweets during Harvey failed to provide sufficient information for life-saving operations, such as locations, special needs, rescue-related hashtags, and first responders' accounts. Second, the spatial-temporal analysis of rescue-request tweets reveals that around 70% of the rescue requests were posted on August 27 and August 28, two days after Harvey made the first landfall (August 25). Most of the rescue-request tweets were sent from three neighborhoods in Houston and one neighborhood in Port Arthur. Over forty percent of these rescue requests were sent by people living in the FEMA-defined low flood risk areas, which could be a result of either inaccurate flood zone map or residents caught off-guard and did not evacuate. Third, we examined the geographical and socioeconomic status of communities where people sent rescue requests and found that most of them had below-average elevations, received more rainfall from Harvey, and had more socially vulnerable populations. Finally, we identified the challenges of using social media for rescue, including users' lack of knowledge on how and who to ask for help online and the need for tools to collect and process rescue requests on social media efficiently.

Several implications arise from this study. First, this research demonstrated that mining social media data is a feasible approach to understand human behaviors. Compared with traditional surveys, which are frequently used in collecting human-centric data for disaster management, social media data provide a fast lens to investigate people's real-time behaviors during hazard events. Second, the identified challenges and recommended improvements of using social media for rescue could guide community preparedness and responses to future disasters. Agencies and organizations are suggested to open social media channels for rescue during large-scale emergencies, spread effective online rescue-request methods to social media users in the mitigation and preparedness phases, and develop tools to collect rescue requests from social media platforms automatically. Users could compose help requests through the recommended strategies when encountering disasters. Finally, first responders could use the vulnerable communities identified in this study to inform emergency management. This study has generated a collection of addresses and block groups where people needed additional evacuation assistance during Hurricane Harvey. First responders could target those areas as vulnerable regions and allocate more rescue resources in future events.

**References**

Blake, Eric S., and David A. Zelinsky. 2018. *Hurricane Harvey*. AL092017. National Hurricane Center Tropical Cyclone Report. National Hurricane Center. https://www.nhc.noaa.gov/data/tcr/AL092017_Harvey.pdf.

Cai, Heng, Nina S. N. Lam, Lei Zou, and Yi Qiang. 2018. "Modeling the Dynamics of Community Resilience to Coastal Hazards Using a Bayesian Network." *Annals of the American Association of Geographers* 108 (5). Taylor & Francis: 1260–1279. doi:10.1080/24694452.2017.1421896.

Cai, Heng, Nina S.-N. Lam, Lei Zou, Yi Qiang, and Kenan Li. 2016. "Assessing Community Resilience to Coastal Hazards in the Lower Mississippi River Basin." *Water* 8 (2). Multidisciplinary Digital Publishing Institute: 46. doi:10.3390/w8020046.

Cutter, Susan L., Kevin D. Ash, and Christopher T. Emrich. 2014. "The Geographies of Community Disaster Resilience." *Global Environmental Change* 29 (November): 65–77. doi:10.1016/j.gloenvcha.2014.08.005.

Cutter, Susan L., Christopher G. Burton, and Christopher T. Emrich. 2010. "Disaster Resilience Indicators for Benchmarking Baseline Conditions." *Journal of Homeland Security and Emergency Management* 7 (1). De Gruyter. doi:10.2202/1547-7355.1732.

Cutter, Susan L., and Christina Finch. 2008. "Temporal and Spatial Changes in Social Vulnerability to Natural Hazards." *Proceedings of the National Academy of Sciences* 105 (7). National Academy of Sciences: 2301–2306. doi:10.1073/pnas.0710375105.

Devlin, Jacob, Ming-Wei Chang, Kenton Lee, and Kristina Toutanova. 2019. "BERT: Pre-Training of Deep Bidirectional Transformers for Language Understanding." *ArXiv:1810.04805 [Cs]*, May. http://arxiv.org/abs/1810.04805.

Dufty, Neil. 2012. "Using Social Media to Build Community Disaster Resilience." *The Australian Journal of Emergency Management* 27 (1): 6.

Federal Emergency Management Agency (FEMA). 2021. "Flood Zones | FEMA.Gov." https://www.fema.gov/glossary/flood-zones.

Gallagher, J.J. 2017. "Hurricane Harvey Wreaks Historic Devastation: By the Numbers." *ABC News*. https://abcnews.go.com/US/hurricane-harvey-wreaks-historic-devastation-numbers/story?id=49529063.

Gomez, Luis. 2017. "Hurricane Harvey: 5 Ways Social Media Saved Lives in Texas." *San Diego Union-Tribune*. https://www.sandiegouniontribune.com/opinion/the-conversation/sd-hurricane-harvey-5-ways-social-media-helped-rescue-efforts-20170828-htmlstory.html.

Guan, Xiangyang, and Cynthia Chen. 2014. "Using Social Media Data to Understand and Assess Disasters." *Natural Hazards* 74 (2): 837–850. doi:10.1007/s11069-014-1217-1.

Houston, J. Brian, Joshua Hawthorne, Mildred F. Perreault, Eun Hae Park, Marlo Goldstein Hode, Michael R. Halliwell, Sarah E. Turner McGowen, et al. 2015. "Social Media and Disasters: A Functional Framework for Social Media Use in Disaster Planning, Response, and Research." *Disasters* 39 (1): 1–22. doi:https://doi.org/10.1111/disa.12092.

Huang, Xiao, Cuizhen Wang, Zhenlong Li, and Huan Ning. 2019. "A Visual–Textual Fused Approach to Automated Tagging of Flood-Related Tweets during a Flood Event." *International Journal of Digital Earth* 12 (11). Taylor & Francis: 1248–1264. doi:10.1080/17538947.2018.1523956.

Jamali, Mehdi, Ali Nejat, Souparno Ghosh, Fang Jin, and Guofeng Cao. 2019. "Social Media Data and Post-Disaster Recovery." *International Journal of Information Management* 44 (February): 25–37. doi:10.1016/j.ijinfomgt.2018.09.005.

Kent, Joshua D., and Henry T. Capello. 2013. "Spatial Patterns and Demographic Indicators of Effective Social Media Content during TheHorsethief Canyon Fire of 2012." *Cartography and Geographic Information Science* 40 (2). Taylor & Francis: 78–89. doi:10.1080/15230406.2013.776727.

Kirby, Ryan H., Margaret A. Reams, Nina S. N. Lam, Lei Zou, Gerben G. J. Dekker, and D. Q. P. Fundter. 2019. "Assessing Social Vulnerability to Flood Hazards in the Dutch Province of Zeeland." *International Journal of Disaster Risk Science* 10 (2): 233–243. doi:10.1007/s13753-019-0222-0.

Kryvasheyeu, Yury, Haohui Chen, Nick Obradovich, Esteban Moro, Pascal Van Hentenryck, James Fowler, and Manuel Cebrian. 2016. "Rapid Assessment of Disaster Damage Using Social Media Activity." *Science Advances* 2 (3). American Association for the Advancement of Science: e1500779. doi:10.1126/sciadv.1500779.

Li, Linna, Michael F. Goodchild, and Bo Xu. 2013. "Spatial, Temporal, and Socioeconomic Patterns in the Use of Twitter and Flickr." *Cartography and Geographic Information Science* 40 (2). Taylor & Francis: 61–77. doi:10.1080/15230406.2013.777139.

Li, Zhenlong, Cuizhen Wang, Christopher T. Emrich, and Diansheng Guo. 2018. "A Novel Approach to Leveraging Social Media for Rapid Flood Mapping: A Case Study of the 2015 South Carolina Floods." *Cartography and Geographic Information Science* 45 (2). Taylor & Francis: 97–110. doi:10.1080/15230406.2016.1271356.

Maldonado, M., D. Alulema, D. Morocho, and M. Proaño. 2016. "System for Monitoring Natural Disasters Using Natural Language Processing in the Social Network Twitter." In *2016 IEEE International Carnahan Conference on Security Technology (ICCST)*, 1–6. doi:10.1109/CCST.2016.7815686.

Mihunov, Volodymyr V., Nina S. N. Lam, Lei Zou, Zheye Wang, and Kejin Wang. 2020. "Use of Twitter in Disaster Rescue: Lessons Learned from Hurricane Harvey." *International Journal of Digital Earth* 13 (12). Taylor & Francis: 1454–1466. doi:10.1080/17538947.2020.1729879.

Morris, David. 2016. "How the 'Cajun Navy' Is Using Tech To Rescue Flood Victims in Louisiana." *Fortune*. https://fortune.com/2016/08/20/cajun-navy-tech-louisiana-flood/.

Qiang, Yi, Nina S. N. Lam, Heng Cai, and Lei Zou. 2017. "Changes in Exposure to Flood Hazards in the United States." *Annals of the American Association of Geographers* 107 (6). Taylor & Francis: 1332–1350. doi:10.1080/24694452.2017.1320214.

Rhodan, Maya. 2017. "Hurricane Harvey: The U.S.'s First Social Media Storm | Time." https://time.com/4921961/hurricane-harvey-twitter-facebook-social-media/.

Shan, Siqing, Feng Zhao, Yigang Wei, and Mengni Liu. 2019. "Disaster Management 2.0: A Real-Time Disaster Damage Assessment Model Based on Mobile Social Media Data—A Case Study of Weibo (Chinese Twitter)." *Safety Science* 115 (June): 393–413. doi:10.1016/j.ssci.2019.02.029.

Sutton, Jeannette, C. Ben Gibson, Nolan Edward Phillips, Emma S. Spiro, Cedar League, Britta Johnson, Sean M. Fitzhugh, and Carter T. Butts. 2015. "A Cross-Hazard Analysis of Terse Message Retransmission on Twitter." *Proceedings of the National Academy of Sciences* 112 (48). National Academy of Sciences: 14793–14798. doi:10.1073/pnas.1508916112.

Sutton, Jeannette, Emma S. Spiro, Britta Johnson, Sean Fitzhugh, Ben Gibson, and Carter T. Butts. 2014. "Warning Tweets: Serial Transmission of Messages during the Warning Phase of a Disaster Event." *Information, Communication & Society* 17 (6). Routledge: 765–787. doi:10.1080/1369118X.2013.862561.

Van Zandt, Shannon, Walter Gillis Peacock, Dustin W. Henry, Himanshu Grover, Wesley E. Highfield, and Samuel D. Brody. 2012. "Mapping Social Vulnerability to Enhance Housing and Neighborhood Resilience." *Housing Policy Debate* 22 (1). Routledge: 29–55. doi:10.1080/10511482.2011.624528.

Verma, Sudha, Sarah Vieweg, William Corvey, Leysia Palen, James Martin, Martha Palmer, Aaron Schram, and Kenneth Anderson. 2011. "Natural Language Processing to the Rescue? Extracting 'Situational Awareness' Tweets During Mass Emergency." *Proceedings of the International AAAI Conference on Web and Social Media* 5 (1). https://ojs.aaai.org/index.php/ICWSM/article/view/14119.

Wang, Jimin, Yingjie Hu, and Kenneth Joseph. 2020. "NeuroTPR: A Neuro-Net Toponym Recognition Model for Extracting Locations from Social Media Messages." *Transactions in GIS* 24 (3): 719–735. doi:https://doi.org/10.1111/tgis.12627.

Wang, Kejin, Nina S. N. Lam, Lei Zou, and Volodymyr Mihunov. 2021. "Twitter Use in Hurricane Isaac and Its Implications for Disaster Resilience." *ISPRS International Journal of Geo-Information* 10 (3). Multidisciplinary Digital Publishing Institute: 116. doi:10.3390/ijgi10030116.

Wang, Zheye, Nina S. N. Lam, Nick Obradovich, and Xinyue Ye. 2019. "Are Vulnerable Communities Digitally Left behind in Social Responses to Natural Disasters? An Evidence from Hurricane Sandy with Twitter Data." *Applied Geography* 108 (July): 1–8. doi:10.1016/j.apgeog.2019.05.001.

Wang, Zheye, and Xinyue Ye. 2018. "Social Media Analytics for Natural Disaster Management." *International Journal of Geographical Information Science* 32 (1). Taylor & Francis: 49–72. doi:10.1080/13658816.2017.1367003.

Wang, Zheye, and Xinyue Ye. 2019. "Space, Time, and Situational Awareness in Natural Hazards: A Case Study of Hurricane Sandy with Social Media Data." *Cartography and Geographic Information Science* 46 (4). Taylor & Francis: 334–346. doi:10.1080/15230406.2018.1483740.

Wu, Desheng, and Yiwen Cui. 2018. "Disaster Early Warning and Damage Assessment Analysis Using Social Media Data and Geo-Location Information." *Decision Support Systems* 111 (July): 48–59. doi:10.1016/j.dss.2018.04.005.

Yang, Z., L. H. Nguyen, J. Stuve, G. Cao, and F. Jin. 2017. "Harvey Flooding Rescue in Social Media." In *2017 IEEE International Conference on Big Data (Big Data)*, 2177–2185. doi:10.1109/BigData.2017.8258166.

Zhai, Wei, Zhong-Ren Peng, and Faxi Yuan. 2020. "Examine the Effects of Neighborhood Equity on Disaster Situational Awareness: Harness Machine Learning and Geotagged Twitter Data." *International Journal of Disaster Risk Reduction* 48 (September): 101611. doi:10.1016/j.ijdrr.2020.101611.

Zou, Lei, Nina S. N. Lam, Heng Cai, and Yi Qiang. 2018. "Mining Twitter Data for Improved Understanding of Disaster Resilience." *Annals of the American Association of Geographers* 108 (5). Taylor & Francis: 1422–1441. doi:10.1080/24694452.2017.1421897.

Zou, Lei, Nina S. N. Lam, Shayan Shams, Heng Cai, Michelle A. Meyer, Seungwon Yang, Kisung Lee, Seung-Jong Park, and Margaret A. Reams. 2019. "Social and Geographical Disparities in Twitter Use during Hurricane Harvey." *International Journal of Digital Earth* 12 (11). Taylor & Francis: 1300–1318. doi:10.1080/17538947.2018.1545878.